\documentclass[manuscript,screen,nonacm]{acmart}
\usepackage{natbib}
\AtBeginDocument{%
  \providecommand\BibTeX{{%
    \normalfont B\kern-0.5em{\scshape i\kern-0.25em b}\kern-0.8em\TeX}}}

\usepackage{geometry}
\usepackage{caption}
\setcopyright{acmcopyright}
\copyrightyear{2018}
\acmYear{2018}
\acmDOI{XXXXXXX.XXXXXXX}

\acmConference[Conference acronym 'XX]{Make sure to enter the correct
  conference title from your rights confirmation emai}{June 03--05,
  2018}{Woodstock, NY}
\acmBooktitle{Woodstock '18: ACM Symposium on Neural Gaze Detection,
 June 03--05, 2018, Woodstock, NY} 
\acmPrice{15.00}
\acmISBN{978-1-4503-XXXX-X/18/06}

\begin{document}
%\addtolength{\evensidemargin}{0in}
\setlength\oddsidemargin{\dimexpr(\paperwidth-\textwidth)/2 - 1in\relax}
\setlength\evensidemargin{\oddsidemargin}
\title{Subjective visualization experiences: impact of visual design and experimental design}

%Reflections on experimental methods to study subjective visualization experiences
 \author{Laura Koesten}
\affiliation{%
  \institution{University of Vienna}
  \streetaddress{TBD}
  \city{Vienna}
  \country{Austria}}
\email{larst@affiliation.org}

\author{Drew Dimmery}
\affiliation{%
  \institution{University of Vienna}
  \streetaddress{1 Th{\o}rv{\"a}ld Circle}
  \city{Vienna}
  \country{Austria}}
\email{larst@affiliation.org}

\author{Michael Gleicher}
\affiliation{%
  \institution{University of Wisconsin - Madison}
  \city{Madison}
  \country{USA}
}

\author{Torsten Möller}
\affiliation{%
 \institution{University of Vienna}
 \streetaddress{tbd}
 \city{Vienna}
 \country{Austria}}

\renewcommand{\shortauthors}{Koesten, et al.}

\begin{abstract}
In contrast to objectively measurable aspects (such as accuracy, reading speed, or memorability), the subjective experience of visualizations has only recently gained importance, and we have less experience how to measure it. We explore how subjective experience is affected by chart design using multiple experimental methods. We measure the effects of changes in color, orientation, and source annotation on the perceived readability and trustworthiness of simple bar charts. Three different experimental designs (single image rating, forced choice comparison, and semi-structured interviews) provide similar but different results. We find that these subjective experiences are different from what prior work on objective dimensions would predict. Seemingly inconsequential choices, like orientation, have large effects for some methods, indicating that study design alters decision-making strategies. Next to insights into the effect of chart design, we provide methodological insights, such as a suggested need to carefully isolate individual elements in charts to study subjective experiences.

\end{abstract}

\begin{CCSXML}
<ccs2012>
   <concept>
       <concept_id>10003120.10003145.10011770</concept_id>
       <concept_desc>Human-centered computing~Visualization design and evaluation methods</concept_desc>
       <concept_significance>500</concept_significance>
       </concept>
   <concept>
       <concept_id>10003120.10003145.10011769</concept_id>
       <concept_desc>Human-centered computing~Empirical studies in visualization</concept_desc>
       <concept_significance>500</concept_significance>
       </concept>
 </ccs2012>
\end{CCSXML}

\ccsdesc[500]{Human-centered computing~Visualization design and evaluation methods}
\ccsdesc[500]{Human-centered computing~Empirical studies in visualization}

\keywords{Data visualization, subjective experience, chart evaluation, evaluation methods}


\maketitle

\section{Introduction}

%Different groups of people widely use charts. 
Charts are widely used by different people. They are increasingly important in public messaging and are shared and discussed online~\citep{DBLP:conf/chi/ZhangSPBBP21}. The use of charts in everyday situations (e.g., on social media or newspapers) poses distinct challenges to visualization design~\cite{kennedy2018feeling}: subjective factors are important for how people react to them.

\emph{Objective} performance of charts, the measure of how charts effectively communicate to the viewer, is well studied~\citep{franconeri2021science}. 
The literature includes a large array of studies exploring the relationship between designs and performance, measured in the context of tasks such as precise reading, decision-making, inferring, or recall~\citep{padilla2018decision,alhadad2018visualizing,cleveland1984graphical}. Beyond traditional evaluation criteria, authors have measured effects on memorability, bias, or cognitive load~\citep{DBLP:journals/tvcg/BorkinVBISOP13,dimara2018task,calero2018studying}.
This has led to guidance on how to design charts from researchers, practitioners, and tool vendors (e.g.,~\citep{DBLP:books/daglib/0034520}). 

However, much less is known about viewers' \emph{subjective} experience of charts. Such responses are important: if the viewer does not \textit{trust} a chart---or dismisses it as not \textit{readable}---the communicative value of the chart is lost, no matter how carefully it might be crafted or displayed. 
Design choices impact subjective experiences of charts~\citep{wang2019emotional,dwyer2021struggling,kennedy2018feeling}. However, they may not impact them in the same ways as they impact objective performance~\citep{anderson2019visual}. 

In this paper, we explore how chart design decisions affect the subjective responses of viewers. In particular, we explore how design decisions that have little expected effect on objective performance can have measurable effects on subjective response. Our study uses very basic charts (simple bar charts) and design decisions (color, chart orientation, and source annotation). While these design variables are not expected to have much impact on the actual objective performance of chart reading, we find they have a substantial impact on subjective experiences. Indeed, we find that these seemingly inconsequential choices can have a greater impact than choices that can make an objective difference. For example, objectively, the horizontal or vertical orientation of a chart should have no effect on whether the chart is trustworthy. Still, the presence of a source annotation should objectively add trustworthiness. Yet, we find orientation can have a larger effect on \emph{perceived} trustworthiness than source annotation.

Unlike objective performance, where there is a strong literature and tradition of empirical methods, less is known about experimental design for understanding subjective responses. Therefore, in this paper, we apply multiple experimental designs to the same question of the impacts of design changes on simple charts. Specifically, we use single-image ranking, forced choice comparison, and semi-structured interviews. While the methods provide similar high-level results, they differ in effect sizes and details, at times strikingly so. Combined, the analysis allows us to bring the three studies together to tell a coherent story about our specific question but also to shed light on methodology for how to study the subjective response to charts.

Our contributions are along three lines:
\begin{itemize}
    \item Visualization research: We add to the state of the art in data visualization by providing insights into the effects of chart design on people's perceived experiences of charts. 
    Our work exposes a difference between how design factors influence objective and subjective responses to charts. 
    This allows us to scope further research in the field and compare and contrast people's perceptions of readability and trustworthiness with existing knowledge from perceptual studies. 
    \item Methodological: We provide insights into methodological questions concerning study designs targeting subjective experience in the context of visualization research, discussing single image ratings and forced choice designs. We show how seemingly minor choices in design can lead to meaningful differences in interpretation and how bringing multiple designs together can lead to more robust models of how charts are perceived.
    \item Understanding people: In-depth accounts of participants' reasoning about the perceived effectiveness of charts beyond dimensions commonly considered in perceptual visualization research.
\end{itemize}

\section{Related work}
In this section, we review typical dimensions and methods used in the study of chart evaluations. We focus on the role that subjective experience plays in evaluating charts, particularly perceived readability and trustworthiness.

\subsection{Chart evaluation}
\label{sec:evalution}

There are many guidelines for how to design charts for certain purposes. These stem from design practice, as well as a long history of empirical studies and theoretical frameworks. Some parts of this vast research space have been more thoroughly explored than others. Much of the research in chart evaluation aims to explain how people process visual representations of data. Numerous studies have investigated perceptual effectiveness in great detail, spearheaded by seminal work such as \citet{cleveland1984graphical} and come with established methodologies developed over the decades~\citep{franconeri2021science}. From these studies, we know that people's perceptions of data are influenced by its visual encoding through marks and channels (such as color, size, shape, etc.)~\citep{DBLP:conf/visualization/Healey96,DBLP:journals/tvcg/HarozW12,DBLP:books/daglib/0025540}. Many studies have shown that chart type and their design variations impact their efficacy for certain tasks~\citep{DBLP:journals/corr/abs-2107-07477}. What these studies have in common is their focus on ``objectively'' measurable dimensions, often perceptual dimensions, which are believed to be identical across a population and independent of individual experiences or knowledge (such as accurately decoding the data). 
In contrast, some of the decisions of data visualization designers are often based on a rule of thumb \citep{parsons2021understanding}, usually less specific, and based on expertise and intuition related to expectations of communicative value and engagement, arguably subjective criteria. This work therefore focuses on subjective evaluations of the chart as a whole by participants. %Because it is unclear what determines these subjective experiences, we used focus groups to inform the concepts we investigate in this study.

For more than a decade, the visualization community has used online methods, including crowdsourcing, to evaluate the effectiveness of visual representations~\citep{van2008perceptual}. 
This is commonly determined using controlled experiments, in which participants are asked to complete tasks depending on the desired dimension of measurement; for a review, see, for 
instance, \citet{DBLP:journals/tvcg/0001ICSM13}. 
Though traditional lab studies prevail when there is demand for greater control~\citep{DBLP:conf/chi/HeerB10,DBLP:conf/beliv/KosaraZ10,DBLP:conf/beliv/Kosara16}, online experimentation is seen as a method, which helps with scale, speed, and diversity of participants~\citep{DBLP:conf/dagstuhl/BorgoLBFJKKMMLB15,DBLP:journals/tochi/YangLZ14,DBLP:journals/cgf/BorgoMBML18,DBLP:conf/dagstuhl/EdwardsKFCC15,DBLP:conf/chi/HeerB10}. At the same time, online tasks are not without their challenges. 

We argue that many factors influencing both ratings and forced choice experiments in visualization evaluation are not captured in traditional online studies, and rarely is qualitative data used to contextualize the results in accounts of participants' thought processes. Critiques surrounding issues with replication have been voiced repeatedly and across disciplines~\citep{paritosh12,baker2016reproducibility}. Not least, therefore, we investigate the influence of minor visual design changes (color, orientation, and source annotation) on subjective visualization experience in-depth, reporting on three methodologically different studies. 

\subsection{Subjective experience of charts}
\label{sec:subjective}
When evaluating charts, experimental studies use dimensions such as precise reading, decision-making, inferring, or recall~\citep{padilla2018decision,alhadad2018visualizing,cleveland1984graphical} but also other dimensions such as long and short term memorability~\citep{DBLP:journals/tvcg/BorkinVBISOP13}, engagement~\citep{DBLP:conf/chi/HungP17}, enjoyment~\citep{DBLP:journals/tvcg/BrehmerM13,DBLP:journals/ivs/SpragueT12}, cognitive load~\citep{DBLP:journals/cgf/AndersonPMSPS11,DBLP:journals/tvcg/YoghourdjianYDL21} or bias \citep{dimara2018task}, usually in task-based settings. %persuasion~\citep{DBLP:journals/tvcg/PandeyMNSB14}
The subjective experience of charts is not often the focus of attention. However, some of the dimensions of measurement from prior studies tend to be more subjective than others, for instance, engagement and enjoyment~\citep{hung2017assessing}. 
%In the data visualization community,
\citet{DBLP:conf/chi/HungP17} evaluate the user's emotional involvement or investment while interacting with a visualization, while Saket et al.~\citep{DBLP:conf/beliv/SaketES16} review evaluation studies that focus on user experience goals. In their survey on data visualisation evaluation, \citet{DBLP:journals/tvcg/0001ICSM13} discuss the user experience of visualisations. This includes evaluations that elicit subjective feedback and opinions on a visualization (tool). While the authors note an increasing interest in user experience studies, data is usually gathered via interviews or questionnaires about a specific system. %Our study aims to collect subjective ratings in a different manner via an experimental online set-up.
%For instance, visual design choices can mediate a persons willingness to engage with digital content %perceptions readability and perceptions of trustworthiness of digital content, such as digital longform articles but also of ML models~\citep{greussing2019simply,gaba2023my,}. 

Other (visual) media are better understood in this regard. 
For instance, previous studies on video recordings, schematic maps, or digit recognition ~\citep{andersen2019visual,bruun2015mind,roberts2017preference} suggest that someone's perceived experience of visual %objects/
stimuli is influenced by a multitude of factors that are not necessarily related to performance measures. This includes, for instance, expectations of what one is likely to see (visual expectations), which can change the quality of the experience without changing performance capabilities associated with that quality~\citep{andersen2019visual}. 
This tension between objectively measurable performance (e.g., memorability, recall, etc., as done in chart perception studies) and user preferences (i.e., subjective experience) is the starting point for our study.

However, subjective responses to visualizations are difficult to study for a variety of reasons~\citep{saket2016beyond,lam2011empirical}, including the absence of a meaningful baseline, a complex interplay of internal and external factors pertaining to participants (such as prior experience, familiarity, literacy, as well as cultural and potentially socioeconomic factors) and an expected, but unspecified, amount of variance. What is more, is that we do not know whether differences in subjective experiences of charts are aligned with objective differences.
Literature from other fields suggests %both a mismatch between subjective experience and performance measures  
subjective experience to be an important factor for engagement (e.g., in the influence of happiness on work engagement \citep{de2020interplay,greussing2019simply}). A mismatch between subjective experience and performance measures is also mentioned in different contexts (e.g., investigating the sense of smell %has poor correlation with actual olfactory ability
\citep{philpott2006comparison}; workload \citep{yeh1988dissociation}, or perceptions of time \citep{hornik1996psychological}). While these are not directly transferrable to the experience of data visualizations, they give reason to assume that subjective experience is a powerful aspect of how we interact with information, also when visually presented. %or the perception of cognitive effects is related more to mood than to objective performance \citep{}) , as well as 
We argue that our understanding of the subjective experience of visualizations is limited to date, particularly concerning questions of readability and trustworthiness.

\subsection{Perceived readability and trustworthiness}
\label{sec:dimensions}
Visual design choices can impact subjective experiences of charts, even at the point of encounter~\citep{DBLP:journals/tvcg/LeeKHLKY16}, and have been shown to impact perceptions of readability and trustworthiness of other digital content~\citep{gaba2023my,greussing2019simply}. 
In this work, we focus on perceived readability and trustworthiness as aspects of the larger concept of subjective experience.  
These %subjective 
dimensions are chosen based on two focus groups as well as on related work, which suggests that they mediate whether a user will interact with, rely and build on the information displayed in a visualization~\citep{DBLP:conf/eurorvvv-ws/MayrHSW19}. 

Both, but especially trust, are multi-dimensional, complex, context-dependent concepts~\citep{wallace2020consuming}. They are, in the context of our study, related to processing fluency, which is the experienced ease of ongoing mental processes in information consumption and decision making~\citep{unkelbach2013general}. Our aim is not to measure these concepts per se but as proxies for two important aspects of chart engagement.%tasks people undertake when interacting with charts. 
Both have been brought up in our focus groups (see \autoref{sec:methodology}), paraphrased here: 
\textit{"I can read and hence understand the chart"}, and as a follow-up \textit{"I believe the chart (and the data) in so far as to make a decision."}. 
Both ease of reading and perceived trustworthiness likely play a key role in engagement with %information 
visualization as a factor that mediates whether a user will rely and build on the information displayed in a visualization~\citep{DBLP:conf/eurorvvv-ws/MayrHSW19}. Usability and user experience (such as a positive look and feel) can increase trust~\citep{DBLP:conf/stast/CostanteHP11}. For instance, \citet{DBLP:journals/topics/JoslynL16} looked at how certain visual elements influence trust in weather predictions, for instance uncertainty representations.
Literature suggests that trust should be assessed in relation to an action rather than in general terms~\citep{dumouchel_2005}. It is a different conceptual task for respondents than direct questions about trust~\citep{oecd2017oecd}. Trust is commonly measured via self-report questions of how much participants trust a visualization or believe in its accuracy, e.g.,~\citep{xiong2019examining}.

\section{Methodology}
\label{sec:methodology}
In this section, we describe charts used across all three studies and the participants as well as their characteristics. We then detail each study design individually, referring to them as Study 1 (\textit{Rating}), Study 2 (\textit{Choice}), and Study 3 (\textit{Qual}).
This work was carried out as part of a project that has undergone ethical screening according to the guidelines of \textit{institution anonymized for review} and has been determined to be low-risk.

\subsection{Input charts}

\begin{figure}
\centering
\begin{minipage}[b]{0.45\linewidth}
\includegraphics[width=0.8\linewidth]{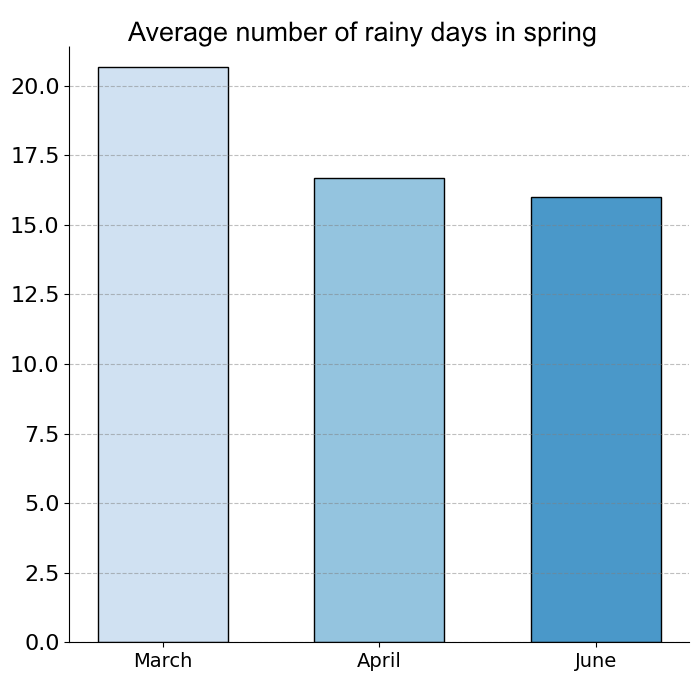}
\captionof{figure}{Example chart used in study}
\label{fig:example_chart}
\end{minipage}
\begin{minipage}[b]{0.45\linewidth}
\def\arraystretch{1.7}
    \begin{tabular}{ p{4cm}l } 
        \textbf{Design parameter} & \textbf{Variations}\\
        \hline
        % only use multirow when the text should be aligned in the middle of the row
        % \multirow{3}{*}{Granularity} & General\\
        %   & Middle detail\\ 
        %   & Specific\\
%        Legend & yes / no \\
        \hline
        \textbf{Source annotation}: fictitious source visible on the chart & yes / no \\
        \hline
        \textbf{Color}: greyscale versus blue 
 & gray / blue \\
        \hline
        \textbf{Orientation}: horizontal or vertical orientation of bars  & vertical / horizontal \\
        \hline
%        Number of data categories  & 2 bars / 3 bars / 15 bars \\
        \hline
    \end{tabular}
    \captionof{table}{Design parameters varied in the input charts}
    \label{tab:design-parameter}
\end{minipage}
\end{figure}

%Fictitious source visible on the chart (publishing institution)
%Grey scale versus blue 
%Horizontal or vertical orientation of rectangles in bar charts

% \begin{table}[h]
%     \centering
%     \bgroup
%     \def\arraystretch{1.7}
%     \begin{tabular}{ ll } 
%         \textbf{Design parameter} & \textbf{Variations}\\
%         \hline
%         % only use multirow when the text should be aligned in the middle of the row
%         % \multirow{3}{*}{Granularity} & General\\
%         %   & Middle detail\\ 
%         %   & Specific\\
% %        Legend & yes / no \\
%         \hline
%         Source annotation & yes / no \\
%         \hline
%         Color & gray / blue \\
%         \hline
%         Orientation & vertical / horizontal \\
%         \hline
% %        Number of data categories  & 2 bars / 3 bars / 15 bars \\
%         \hline
%     \end{tabular}
%     \egroup
%     \caption{Design parameter varied in the input charts}
%     \label{tab:design-parameter}
% \end{table}

The charts were generated automatically based on randomly generated data for temperature, rainfall, and cloudiness. The choice of parameters was informed by historical UK climate data given by the Met Office\footnote{\url{https://www.metoffice.gov.uk/}}. We opted for charts showing fictive weather data because it is a relatively neutral topic that is familiar to most participants. 

We chose basic bar charts as they are one of the most common visualization types, considered to be simple and well-understood and match the data we use. The charts varied across three design parameters (source annotation, color, and orientation), as can be seen in \autoref{tab:design-parameter} and are available in the supplementary material. For color, the charts were either monochrome blue or monochrome grey, and we chose 3 shades such that the smallest bar had the darkest shade.

These variations were chosen because, based on both the visualization literature as well as based on our pilot studies described in \ref{pilots}, they are not seen to objectively alter how well a chart could be read or trusted.
The exception is a likely effect of source annotation on trustworthiness, which was chosen because it was expected to differentiate between the two measured dimensions.

In detail, we hypothesize the design parameters to be inconsequential for participants' subjective ratings of readability and trustworthiness. We would not expect color and orientation to have an objective impact on readability or trustworthiness because they do not change the content of the chart. We would expect source annotations to have a positive impact on trustworthiness because they provide provenance of the data. %These intuitions were confirmed in the pilot study described in \ref{pilots}. 

\subsection{Participants}
Study participants were recruited from students of a Human-Computer Interaction bachelor course. One reason was to allow access to a very similar, but not identical, cohort for all studies, where the qualitative in-person study presents the bottleneck for recruitment. Participation was voluntary and incentivized with a very low level of course credits. \autoref{tab:participants} gives an overview of the participants for each study.

%86 students took part in Study 1 (\textit{Rating}), 127 students took part in Study 2 (\textit{Choice}) and 20 took part in Study 3 (\textit{Qual}).

%After piloting the study design on crowdsourcing platforms, we opted for a slightly more controlled setting for recruiting students. One reason was to allow access to a very similar, but not identical, cohort for all studies, where the qualitative study presents the bottleneck.

\subsection{Scoping pilots}
\label{pilots}
To define the scope of the study, we relied on the authors' experience in data visualization, on related literature, e.g. Quadri and Rosen~\citep{DBLP:journals/corr/abs-2107-07477}, and on insights from two pilot focus groups, aiming to identify design parameters that would influence readability and trustworthiness. From the discussed parameters, we chose one that was most often mentioned to determine trustworthiness (which was source annotation), one that was mentioned to potentially influence perceptions of readability (which was color), and bar orientation as a parameter that did not seem to impact either of the dimensions. We further discussed the wording of the questions used in the experimental tasks as part of these pilots. The focus groups were attended, respectively, by $7$ and $4$ computer science students with practical experience in data visualization (M.Sc./Ph.D. level), all frequent designers of charts. We used individual brainstorming methods, including sketching, as a form of communication and prompts for subsequent group discussion. The groups were both moderated by the same author, attended by two authors, and discussed to reach a consensus on which design parameters to focus on.

\subsection{Question selection}
Targeting the dimensions \textit{readability} and \textit{trustworthiness}, we tested several question wordings in the pilots as well as in discussions with two professors of psychology. Readability was prompted via the question of whether the chart was perceived to be easy to explain to another person. This was chosen informed by the concept ``ease of use''~\citep{karahanna1999psychological}, which was best understood in our pilots as ``easy to read''. When directly asked whether a chart was considered easy to read, some pilot participants interpreted the questions as a simple reading task, focusing on the labels without thinking about the message of the chart. Hence we related the question to the imagined action of being able to explain the chart to another person, as was recommended also by one of the psychology experts consulted in the study design phase, as well as in the literature.

\subsection{Task description and design}
We chose three commonly used methods in visualization evaluations which can be used when subjective aspects are taken into account: single image rating, forced choice, and a qualitative approach.
Participants were shown the chart(s) and the corresponding question on the same screen to capture instantaneous responses while engaging with the chart, which may be different than retrospection~\citep{bruun2015mind}. All participants were asked pre-task questions regarding demographics and visualization experience. All studies were additionally piloted with a minimum of three participants. Chart order, as well as question order, was randomized throughout all tasks. 

\paragraph{Study 1: Single image rating task}
The study was conducted online. After seeing a chart, participants were asked to rate the chart on a 5-point Likert scale according to two dimensions: the ease with which they were able to read the chart and how trustworthy they found the chart. In a between-subjects design, respondents were shown eight charts, varied across the three parameters described in \autoref{tab:design-parameter}, and asked to rate them individually.
%How easy would it be for you to explain this to a friend ( very hard / very easy) %5 points (1=very easy, 5=very hard)
%Do you consider this chart trustworthy? (not at all / very much) %5 points (1=very much, 5=not at all)
%Each of these questions was an 5-point Likert scale.

\paragraph{Study 2: Forced choice task}
The study was conducted online. Rather than rating a chart, subjects were presented with two charts that differed in one of the three parameters described in \autoref{tab:design-parameter} and asked to choose which chart was preferred for readability and trustworthiness, respectively. %each of three dimensions: readability and trustworthiness as before, but also explainability. %which of the charts was considered easier to explain to a friend.
Participants were presented with eight comparisons. The charts were displayed on top of each other on the same screen, in randomized order, and with a large label (A/B) on the image, which was used to choose the better chart on a four-point scale.
Analyses collapsed this scale to a binary measure of which chart was preferred (our analysis shows that results did not change substantively due to this coding choice). This binary scale provides better interpretability in our subsequent analyses.
%Please rate which of the charts better fits the criteria:
%Easier to read
%More trustworthy
%Easier to explain to a friend

\paragraph{Study 3: Semi-structured interviews and think-aloud task}
The study was conducted in person, concurrently with conducting an online task as a discussion prompt, and took around 30 minutes. Participants were asked to think aloud while being presented with four forced-choice chart comparisons from Study 2, as well as additional real-world charts used as discussion prompts, which were modified slightly according to the design parameters in \autoref{tab:design-parameter}. These were two bar charts published by Statista consumer insights\footnote{\url{https://de.statista.com/}}, an online platform offering statistics and reports. The charts were not embedded in the study directly but simply used as discussion prompts to contextualize the findings from the experimental task.
%https://de.statista.com/infografik/23357/anteil-der-befragten-die-diese-online-bezahldienste-nutzen/
% https://www.statista.com/chart/27584/how-common-are-online-returns-gcs/
The online task set-up was identical to Study 2 (\textit{Choice}), but after each comparison, participants were asked to reflect upon the individual charts. In addition to asking about readability and trustworthiness, we also included one additional dimension: "liking a chart". This was added to allow a more in-depth exploration of the discriminatory power of the rating dimensions used in the prior studies.
The study was conducted in German; all quotes were translated and proofread by two of the authors (fluent in both German and English). Participants' ratings were recorded in the online task; the study conductors additionally took notes and interviewed participants.

\subsection{Analysis}

Results are estimated and visualized based on average values of the Likert scale for the outcome of interest. Implicit in this approach is marginalization over all other background factors~\citep{delacuesta2022improving}.
This means that the quantities displayed represent the average change in response from changing a single design parameter (e.g., from horizontal to vertical bars) while randomizing over all other design parameters.
\citet{hainmueller2014causal} provide a rigorous treatment of this approach to the analysis of factorial experiments.
It demonstrates how much the average response would differ in the experiment as that factor's level is varied.
Standard errors are clustered to adjust for the fact that individual responses within a subject are not independent using the estimator of \citet{bell2002bias} with a degree-of-freedom adjustment from \citet{pustejovsky2018small}.
This is necessary because, for example, users may have some characteristic that makes their outcomes similar: like an ``agreeableness'' quality that makes them more likely to respond positively to questions.
For more on the necessity to cluster standard errors in this design, see~\citet{abadie2017should}.

Study 3 (\textit{Qual}) was analyzed via a combination of inductive and deductive coding based on reflexive thematic analysis \citet{braun2023toward} using the qualitative data analysis software ATLAS.ti\footnote{\url{https://atlasti.com/}}. The aim was to identify participants' decision-making criteria by surfacing their thought processes when being presented with charts with differing design parameters. Codes were structured according to the design parameters (orientation, color, source annotation) but also according to the subjective dimensions we studied (readability, trustworthiness, and, in the case of Study 3, also "liking a chart"), which allowed to understand the relation of an emerging theme to these parameters.

\begin{table}[h]
    \centering
    \bgroup
    \def\arraystretch{1.7}
    \begin{tabular}{ llll } 
        & \textbf{Study 1 (\textit{Rating})} & \textbf{Study 2 (\textit{Choice})} & \textbf{Study 3 (\textit{Qual})}\\
        \hline
        \hline
        Participants & 86 & 127 & 20\\
        \hline
 %       Age & xx & xx &\\
 %       \hline
        Taken formal VIS class & 13\% & %21
        19\% & 15\%\\
        \hline
        Semester of study & 5.7 (3.3) & 5.5 (2.3) & 6.0 (2.7)\\
        \hline
        Confidence* in reading charts & \raisebox{-0.5\height}{\includegraphics[height=18pt]{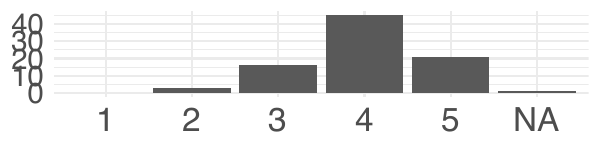}} & \raisebox{-0.5\height}{\includegraphics[height=18pt]{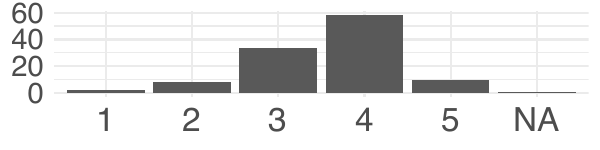}} & \raisebox{-0.5\height}{\includegraphics[height=18pt]{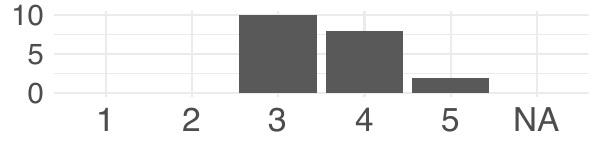}}\\
        \hline
        \hline
    \end{tabular}
    \egroup
    \caption{Descriptive Statistics about participants. Binary variables are provided with the percentage of the sample with that characteristic. Continuous variables are provided as the mean response (with standard deviation in parenthesis). Factors are presented as in-line histograms. (*1=not confident at all, 5=very confident) }
    \label{tab:participants}
\end{table}

\section{Results}

We present the results for each study separately, noting that a direct numerical comparison of the results is not warranted due to the differences in the experimental set-up of the studies.

\subsection{Study 1 (\textit{Rating})}

\begin{figure}[h]
\centering
\includegraphics[width=0.6\linewidth]{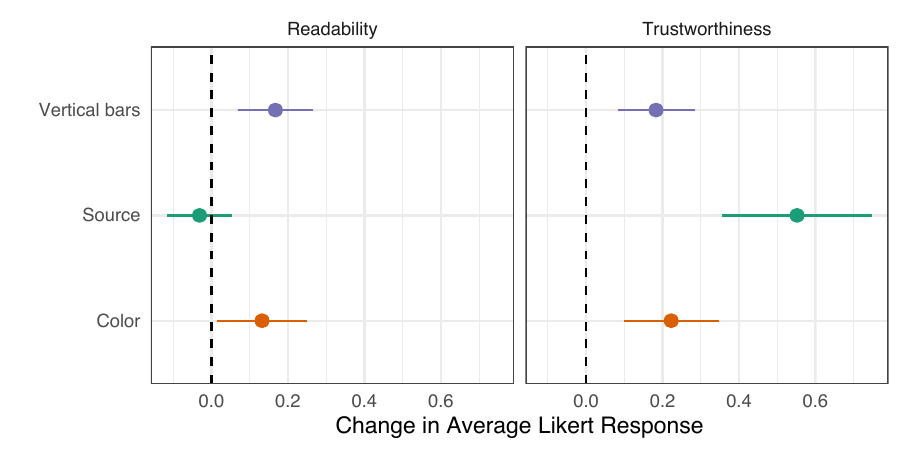}
\caption{Study 1: Average effects over 86 respondents' ratings over 8 charts each (standard errors adjusted for clustering). Respondents tended to rate colorful and vertical charts higher on readability, but source citations only improved ratings for trustworthiness.}
\label{fig:overall_study1}
\end{figure}

\autoref{fig:overall_study1} shows the overall effects of chart design parameters on their subjective rating of the chart. A few points stand out: colored charts are rated as both easier to read and more trustworthy than greyscale charts by 1 to 2 tenths of a point on the Likert scale.
The presence of a source citation on a chart does not markedly change ratings of ease but substantially increases trustworthiness ratings by more than half a point on the scale. Vertical bars are rated as easier and more trustworthy than horizontal bars by around one-fifth of a point on the scale.
No significant heterogeneity in effects was discovered based on whether respondents had taken a visualization class or their confidence in reading charts.

\subsection{Study 2 (\textit{Choice})}

\begin{figure}
    \centering
    \includegraphics[width=0.6\linewidth]{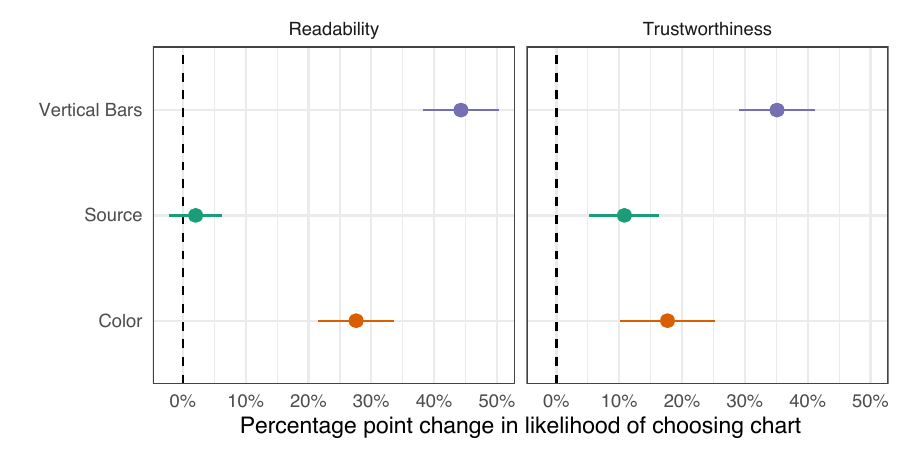}
    \caption{Study 2: Average effects in pairwise comparison preferences over 127 respondents' choice in 8 pairwise comparisons. Respondents tended to prefer vertical and colorful charts on both readability and trustworthiness but preferred source annotation only for trustworthiness. An effect of 20pp for color indicates that a chart with color was preferred by 20pp over charts in greyscale.}
    \label{fig:overall_paired}
\end{figure}

\autoref{fig:overall_paired} shows the average results for the experimental design in which users are asked to choose between two charts. We see that orientation and color have large effects on whether a chart is chosen as more readable and trustworthy. In contrast, source annotation has a smaller impact on trustworthiness and a negligible impact on readability. That is, the choices we would expect to have a negligible objective impact have considerable subjective impact. In contrast, a choice that we expect to have an objective impact has a smaller impact on the subjective responses. 

This is in contrast to Study 1 (\textit{Rating}), in which source annotation was an important determinant of rated trustworthiness.
Nevertheless, more aesthetic and salient design elements had a large impact on preferences.
Respondents strongly preferred vertical over horizontal bars (40-50\% increase in the chance they preferred a given chart) and colorful charts over greyscale (20-30\% increase).
No significant heterogeneity based on subgroups was found (see the supplement).

\autoref{fig:indiv_paired} provides more of a look at individuals' preferences with respect to color and orientation.
The x-axis shows the fraction of the time a respondent chose the color chart when presented with one color and one grayscale chart.
The y-axis shows the same quantity but for the comparison between vertical and horizontal.
Respondents tend to be on the upper half of both of these dimensions: more prefer color to grayscale and vertical to horizontal charts.
This is quantified in the table in Figure~\ref{fig:indiv_paired}.
Moreover, the fact that there are very few respondents in the lower left quadrant indicates that these preferences are related: a respondent who prefers grayscale will essentially never \emph{also} prefer horizontal charts.
On the other hand, respondents who prefer both color and vertical charts are very common.

\begin{figure}
    \begin{minipage}{0.5\linewidth}
    \includegraphics[width=\linewidth]{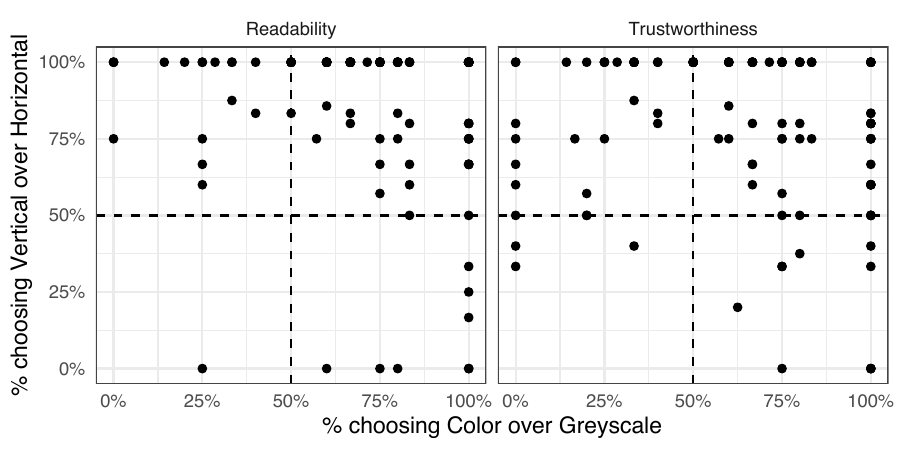}
    \end{minipage}\hfill
    \begin{minipage}{0.45\linewidth}
    \centering
    \emph{Readability}
    \def\arraystretch{1.7}
    \begin{tabular}{r|c|c|}
            & Preferred Greyscale & Preferred Color\\
    \hline
        Preferred Vertical &25 (22\%) & 76 (68\%) \\
        \hline
        Preferred Horizontal &1 (1\%) & 10 (9\%)\\
        \hline
    \end{tabular}
    
    \emph{Trustworthiness}
    
    \begin{tabular}{r|c|c|}
        & Preferred Greyscale & Preferred Color\\
        \hline
        Preferred Vertical & 34 (30\%) & 57 (51\%) \\
        \hline
        Preferred Horizontal & 6 (5\%) & 15 (13\%)\\
        \hline
    \end{tabular}
    \end{minipage}
    
    \caption{Study 2: Individual preferences in pairwise comparisons. The axes show the fraction of respondents, when presented with the choice between a color vs. greyscale chart (or vertical vs. horizontal), who chose the former: each dot is one of the 127 respondents. Individuals consistently preferred colorful and vertical charts.}
    \label{fig:indiv_paired}
\end{figure}

\begin{figure}
    \centering
    \includegraphics[width=0.6\linewidth]{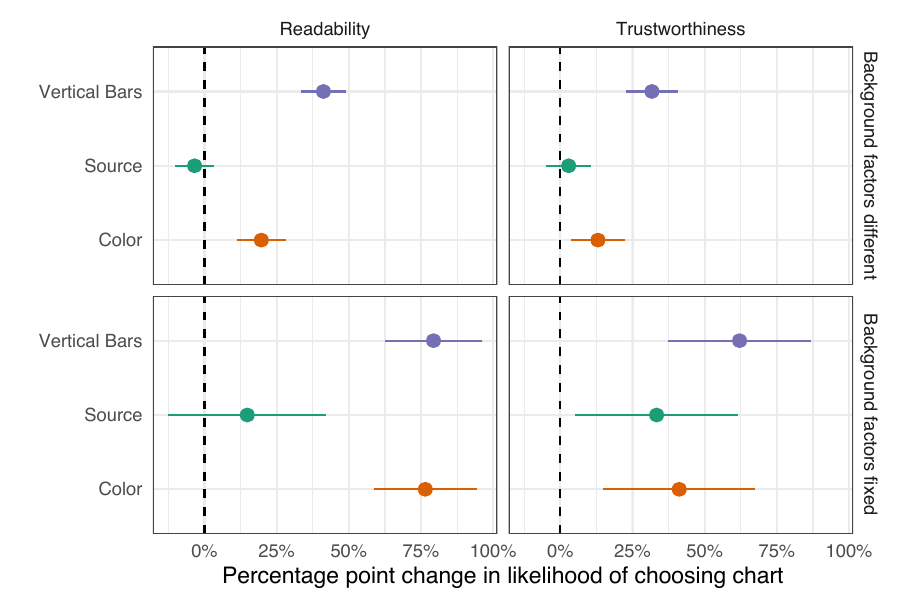}
    \caption{Study 2: Overall effects broken down by whether background factors are the same between choices from 127 respondents with between 1 and 4 comparisons (when background factors are fixed) and between 4 and 8 comparisons (when background factors vary). Effects tended to become larger when background factors were fixed. Only with background factors fixed was an effect of source citation on trustworthiness detectable.}
    \label{fig:paired_by_background}
\end{figure}

\autoref{fig:paired_by_background} demonstrates overall effects when broken down in one of two ways.
The top panel shows effects when background factors are \emph{different}.
For example, it shows the effect of orientation on choice when comparison images differ in source and color.
The bottom panel shows the effect when background factors are held fixed.
For example, the effect of orientation on choice when comparison images look identical in every other respect.
This is a valid comparison because conditioning on a uniformly randomized factor does not induce confounding with respect to other factors.
This comparison, however, demonstrates that background factors strongly effect how a respondent views a chart.
The effect of orientation on their choice of trustworthiness, for instance, is two times larger when charts are otherwise identical than when other factors vary as well.
This pattern is particularly useful when understanding the effects of displaying a source citation.
The overall effects on the choice of chart for trustworthiness are not detectable when background factors vary but are distinguishable from zero only when background factors are held fixed.
This shows that the source annotation has, at best, a minimal effect when there are other aesthetic differences between charts.

%The x-axis shows the fraction of the time a respondent chose the color chart when presented with one color and one grayscale chart.The y-axis shows the same quantity but for the comparison between vertical and horizontal.

\subsection{Study 3: Qualitative Findings}

In this section, we draw on Study 3 (\textit{Qual}) to report on our participants' diverse reasoning for preferring a particular design choice/parameter over its alternative. The aim was to triangulate methods to add context and depth to the results from Studies 1 and 2 and so to offset the limitations of the quantitative approach \citep{bryman2006integrating}.

\subsubsection{Design parameter: orientation}

The strongest preference was articulated for vertical over horizontal bar charts (for 19 out of 20 participants) based on a diversity of reasons. Many participants described a vertical orientation as more pleasant, more intuitive, more common, and familiar, as well as easier to read and explain, or simply as ``better''.

Aspects of readability, such as reading direction, also stood out in participants' observations. 
It was reported to be easier to orient themselves on the x-axis than on the y-axis, meaning that the choice of what variable is displayed on which axis played a role. 
Several participants pointed to reading direction as a reference for decision-making, given that they are used to reading from left to right as opposed to vertically. 

In some participants' accounts, the vertical orientation afforded comparison tasks between the bars better than the horizontal one. 
A participant also mentioned that it was easier to verify the data on the vertical version of the bar charts. 
It was also reported that the numeric labels were perceived to look more tidy on the horizontal axis than on the vertical axis.
A regular connection was made between the visual design and the topic of the chart, which displayed weather data. 
Several participants articulated that a vertical orientation was a logical visual design choice for the number of rainy days:
\begin{quote}
    Especially when it comes to rain, the vertical chart feels more logical. Perhaps because the rain comes from above. Horizontal is less intuitive for this topic. But this is only related to the topic. (P3)
\end{quote}
Interestingly, some participants also connected vertical orientation to perceived ``objectiveness'' and, hence, trustworthiness. One participant even mentioned: 
\begin{quote}
    Despite the lack of source citation, I trust A more [A had no source and a vertical orientation as opposed to a horizontal chart with source annotation] because they use a more classic presentation for the data, more professional. (P15)
\end{quote}

The one reason given by participants for why a horizontal orientation might be preferred was if a time axis was displayed.

\subsubsection{Design parameter: source annotation}

Not all participants picked up on the presence or absence of a source annotation on the charts, which would explain the relatively limited effect size in Study 2. Where the source was noticed, participants mentioned it as a reason to perceive the chart as more trustworthy, sometimes also as the only reason to determine trustworthiness.

For some participants, the source also influenced whether they liked a chart and whether or not they thought it was easy to explain to someone else:  
\begin{quote}
    In a discussion, it is easier to present a chart with a source as an argument because it looks more serious (P8) \textit{or, as put by another participant} is useful in an argumentation. (P11)
\end{quote} 

Several participants expressed liking the chart better when it had a source annotation. One
also mentioned that the presence of a source annotation ``makes it easier for them to read data''. 

\subsubsection{Design parameter: color}

Comparing a grayscale and a colored (monochrome blue) version of the bar chart, most positive attributions were made for the color condition. However, the concepts mentioned by participants for color versus gray differed. Colored bar charts were reported to be more pleasant, refreshing, friendly, aesthetic, and less official. Color was mentioned to add structure and hence improved perceptions of readability.
Color was also connected to memorability and explainability:
\begin{quote}
    Charts with color are easier to remember and, therefore, easier to explain. (P6)
\end{quote}

Whether or not a chart was ``liked'' the most was also seen to be influenced by color. Some participants also made a connection to readability here, being potentially positively influenced by color and hence more likable.

In summary, the majority of participants perceived colored charts as being easier to explain, read, and like but also as having the potential to be perceived as more ``persuasive'', potentially in a negative context. 
Color was, in this context, mentioned to influence trustworthiness ratings. However, participants were split as to whether or not color or grayscale was more influential in their decision-making:
 \begin{quote}
     However, the graphic with color looks more professional and, therefore, potentially more trustworthy. (P11)
 \end{quote}
Gray, on the other hand, was more often described to impact perceptions of trustworthiness by looking more professional, not distracting, more respectable, serious, and connected to print media such as newspapers as well as scientific publications. Gray was considered as not ``screaming for attention'' or obviously ``aiming to persuade'', but also as ``boring''.

Participants also connected the choice of color to the topic:
\begin{quote}
    The blue color suits the rainy days. Therefore, it is more intuitive. [..] (P1)
    \newline
    The blue color best describes the facts: (rainy days) in this case. (P10)
\end{quote}
 
\subsubsection{Question interpretation} 

Here, we present insights into participants' interpretation of the questions posed in the task. While the experiments refer to perceived readability and trustworthiness, we expanded the dimensions for Study 3 to include four concepts: i. easy to read, ii. easy to explain, iii. trustworthy, and iv. liking a chart. 

The first two were chosen to get a deeper understanding of readability, a question that was alluded to in our pilot studies. We found that while these were interpreted slightly differently, they were rated almost identically. Participants' justification for a chart being easier to explain matched more closely with our understanding of readability. This finding confirms the pilot studies where ``easy to read" was at times rated very highly simply when text or values on the chart were considered readable but did not result in a deeper engagement with the message of a chart.

We included liking a chart as a dimension for validation purposes to get insights into the impact of untargeted/unjustified preferences on the results of Study 1 (\textit{Rating}) and 2 (\textit{Choice}).

The following dimensions have been mentioned to contribute to the participant's perception of \textbf{trustworthiness}: source annotation, context about data collection, gray color (``looks more professional, one is not distracted''; ``looks more corporate'').

Other aspects that influenced perceptions of trustworthiness were familiarity with the chart type and their vertical orientation from news sources, even to the extent that the horizontal one looks ``wrong". 

When only the orientation differed between the charts, some participants, therefore, chose the vertical version to have greater trustworthiness. 
Some participants mentioned the manipulative properties of color, which obviously ``aim to persuade". 

\noindent \textbf{Readability} was, as motivated in \autoref{sec:methodology}, prompted via two separate questions, one targeting the perceived ``ease of use'' of the chart and one related to the imagined action of being able to explain the chart to another person. 

As mentioned, the concept was connected both to the orientation of the chart (particularly, the reading direction was called out here) but also to aspects of color: 
\begin{quote}Because it is more comfortable for the eyes. The contrast is also better with the blue colors than with gray. (P5)
\end{quote}

Readability was further attributed for reasons of appropriateness of the used scale to the range of values, and connected to engagement and preference, for instance:
\begin{quote}[..] I'd prefer to look at it. Then, I also explain it better. It looks friendlier. (P2)
\end{quote}

The question was by some interpreted as a persuasive activity, which becomes apparent in responses such as:
``it is more credible with source'' (P12), or 
``can be used as argumentation in a discussion'' (P8).

\noindent \textbf{Liking} a chart was associated with a diversity of factors with a focus on aesthetics, connected to the presence of color, but also with a source annotation, as well as with natural or common orientation (vertical), as this was perceived to be easier to read. While participants were able to give clear reasons for ``liking'' a chart, it was evident that the concept was clearly approached differently to readability and trustworthiness. The think-aloud task showed that each dimension was reasoned for individually, even if they might reach the same conclusions (which can partially be attributed to the limited changes in visual design and the simplicity of the charts).

\noindent \textbf{In summary}, individual participants were very clear about their decision-making criteria and preferences. While the majority mentioned (vertical) orientation as their key dimension for decision-making, some also named the presence of color, and some both. 
Interestingly, these criteria did not necessarily hold when looking at the ``real charts'' where other confounding factors were mentioned or, in the example of color, too many colors were experienced as distracting or as less trustworthy, resembling lower quality popular media, others found the horizontal real chart easier to read in this case.

\section{Discussion}

We believe our results add a layer of understanding of design choices in charts, which complements prior chart perception works that tend to focus on task-dependent, objective measurements of effectiveness or engagement.
This work illustrates how participants' thinking is shaped by 
the choice of the experimental set-up, the formulation of the questions (or measured dimensions), as well as by design changes in input charts.

\subsection{Experiment design}

The primary distinction in Study 1 and 2 designs is whether respondents simply rate charts one at a time or whether they are shown two charts and are forced to choose.
As discussed in \citet{bansak2021conjoint}, these designs may lead respondents to focus on different elements of the charts in making their decisions.
\citet{tversky1988contingent}, in particular, highlight that a difference in measurement task changes the salience of differing features of a task.
In particular, more prominent dimensions tend to be more important to response in choice tasks than in rating tasks.
The idea that differing measurement tasks should yield the same exact results is referred to by \citet{tversky1988contingent} as ``procedure invariance''. In contrast, we find substantial procedural variance.

Given the large design space in composing experiments for the evaluation of data visualizations, our work highlights the impact of small design choices on potential takeaways. The combination of chart type, design choices, questions, and experiment design influences participants' focus as well as effect sizes. %This means that when measuring the effectiveness of visualizations, we likely need to be particularly aware of confounding effects.
This was reflected in findings from Study 3 (\textit{Qual}), where we saw very linear decision-making during the study task, which was adapted in the confrontation with real-world charts where other factors played a role (e.g., the presence of icons). This means that prior work investigating the perceptual effectiveness of visualizations might need to be re-evaluated for confounding factors due to task set-up and design changes in input charts. 

The fact that results are different with a choice task rather than a rating task is consistent with prior work studying factorial designs (in which multiple factors are independently randomized at the same time) as our studies use.
\citet{hainmueller2015validating}, for example, find that the pairwise comparison design better matches the benchmark of how subjects would make decisions in the real world than does assessing single candidates at a time. In their case, the particular task of interest was in evaluating candidates for naturalization in Switzerland.

\citet{evangelidis2023task} demonstrates that, even if decision-making in rating and choice tasks are premised on the same preferences, choice tasks tend to better reflect underlying differences in preferences relative to rating tasks.
In short, the choice task focuses respondents on a more lexicographic process of evaluation. That is, they evaluate individual elements one at a time from most to least important. In contrast, the rating task means that respondents evaluate charts less based on comparative aesthetics, which might lead to the increased effect of source annotation in Study 1 (\textit{Rating}).
Subjective experience needs to be understood within a broader cognitive context, taking into account the multi-dimensionality of influencing factors on experience, which cannot be directly correlated to performance measures~\citep{andersen2019visual}.

Next, we discuss our findings related to each design parameter. It is important to note that due to the differences in study design and analysis, the results across studies are not directly comparable (in terms of effect size) and need to be interpreted in the sense of triangulation and explanation rather than replication~\citep{bryman2006integrating}.

\subsection{Visual design parameters}

We see clearly visible differences concerning subjective ratings of both readability and trustworthiness despite the minor differences in the design parameters of the input charts. Generally speaking, color, vertical orientation, and the presence of a source annotation each impact responses positively. The relative ordering and magnitude of these effects, however, are not always intuitive: vertical orientation is a strong factor in respondents' choices of preferred charts, as is color. Indeed, these small changes in aesthetic design choices result in large changes in respondents' perceptions. We additionally show that people's perceptions of chart elements interact in potentially complex ways and are influenced by the experimental design.   
We now discuss each design parameter individually.
%Source annotation shows a larger effect on trustworthiness ratings in Study 1 (\textit{Rating}). Study 2 (\textit{Choice}), however, shows large effects, especially concerning the positive impact of vertical bar orientation but also of color. Study 3 (\textit{Qual}) added depth and validity to the findings by capturing participants' reasoning activities when making a choice between two visualizations.
%We see a need to consider the influence of differences in the salience of design changes on experimental results. Combined, this suggests a need to carefully isolate individual elements in experimental chart evaluation, even if the changes are seemingly inconsequential. 

\subsubsection{Bar Orientation} 
%\paragraph{Study 1 \textit{Rating}}
Bar orientation impacts both readability and trustworthiness in Study 1 (\textit{Rating}); vertical bars have significantly more positive ratings than horizontal ones. 
%\paragraph{Study 2: Choice}
Effects have the same trend but are strikingly large in Study 2 (\textit{Choice}).

\citet{DBLP:journals/cgf/ConatiCHST14} found working memory to be affected by orientation; users with low visual working memory performed faster with a horizontal layout. Similarly, despite their hypothesis,\citet{fischer2005designing} found vertical bar charts to positively affect the speed of decision-making as opposed to horizontal bar charts, which was partially explained with a suspected higher familiarity with vertical bar charts; a point which was made repeatedly by our participants in Study 3 (\textit{Qual}) and also recently seen in a study about perceived typicality of chart types~\citep{reimann2022typicality}.

The reasons articulated by participants in Study 3 (\textit{Qual}) for choosing vertical over horizontal bar charts are diverse and include aspects of familiarity but also go beyond that. They include questions of reading direction, axis labels, and a preference to "orient" themselves on the x-axis. But they also include a semantic connection to the topic of the chart (e.g., the visual metaphor of the rain coming from above filling up the bar). This thought is mirrored in \citet{setlur2022functional} but limited in empirical evidence to date. These findings add an in-depth understanding of bar chart experiences to the current literature.

\subsubsection{Color} 
In our study, we treated \textbf{color} only in general terms; the chart engagement task was by design under-specified, and the experimental setting did not allow us to target hue, luminance, and other more specific aspects. 
%\paragraph{Study 1: Single chart}
Study 1 (\textit{Rating}) suggests that grayscale charts are perceived to be less readable and trustworthy than monochrome blue ones. 
%\paragraph{Study 2: choice}
Study 2 (\textit{Choice}) also showed a strong preference for monochrome blue charts over grayscale ones, a trend which was mostly reflected in Study 3 (\textit{Qual}). Participants argued particularly about engagement and readability, and they frequently mentioned aesthetic categories.
From a study by Borkin et al.~\citep{DBLP:journals/tvcg/BorkinVBISOP13}, we know color has a positive effect on memorability, which is likely connected to engagement. Colors that are semantically resonant are also known to impact speed on chart reading tasks positively~\citep{lin2013selecting}. Rain, being water, can be associated with blue color, which has been mentioned by some participants in Study 3 (\textit{Qual)}. Considering other topics, this opens questions about the impact of culture on chart experiences as well as age, considering the advance of colored print and digital media within the past decades. However, for trustworthiness ratings, responses were more divided, with arguments for greyscale charts emerging, mostly due to their ``serious" and more ``formal" appeal.

%Add Hypothesis re culture + color: this creates the hypothesis, that this is impacted by culture (that is present in the country you grow up in); age (when you were educated in the 70ies, color was used sparingly ... )
%Color has also been seen to effect memorability positively in a study by Borkin et al.~\citep{DBLP:journals/tvcg/BorkinVBISOP13}.

%expert-chosen semantically-resonant colors improve speed on chart reading tasks compared to a standard palette, an
%Show that response time was positively impacted

% from qual findings: Gray, on the other hand, was more often described to impact perceptions of trustworthiness by looking more professional, not distracting (SJ02), more respectable (SJ06), serious (AS05), and connected to print media such as newspapers (SJ06) as well as scientific publications (AS06b). gray was considered as not "screaming for attention" or obviously "aiming to persuade", but also as "boring" (SJ05).

\subsubsection{Source annotation} 

%\paragraph{Study 1: Single chart} 
In Study 1 (\textit{Rating}), the presence of a \textbf{source}, such as the publishing institution, impacts trust positively but not readability, which matches our assumptions. The context sensitivity of subjective interpretations in data visualizations has been discussed by, e.g., \citet{DBLP:journals/tvcg/KarerHL21}. \citet{li2018communicating} argued that the credibility of a visualization is connected to source attribution in different contexts, and the connection between trust and transparency has been made across disciplines (e.g., \citet{DBLP:conf/hci/VisserCFP14,schnackenberg2016organizational}). This trend was confirmed in Study 3 (\textit{Qual}), where participants mentioned the presence of a source as a reason to perceive a chart as more trustworthy. However, given that source was the design parameter that was least visually salient, some participants did not notice it at all. In contrast, the other parameters (color and bar orientation) present a considerably bigger change. Hence, bar orientation and color were also reported to influence trustworthiness ratings. Another factor that might have impacted the effect size for the trustworthiness dimension was the size of the chart, as displayed in the experiment. Due to the necessity to display two images at the same time, chart size was slightly reduced in contrast to Study 1 (\textit{Rating}); however, this was mitigated by displaying the images on top of each other.

%This matches findings that show the positive effect of transparency on trust~\citep{DBLP:conf/hci/VisserCFP14}. 

\subsubsection{Other influencing factors}
The connection between the visual design and the topic displayed was an apparent factor in our participants' reasoning, both related to color as well as to orientation. This points towards the importance of careful contextual visual design.

It would be interesting to investigate whether participants' perception of the frequency of vertical bar charts in news sources holds true, also across different countries. The effect of familiarity is well known and discussed in-depth in literature in that it predisposes a person toward the stimulus when it is encountered at a later time~\citep{janiszewski2001effects}.

Our findings, if generalizable, have implications beyond the direct contributions we describe in this work.
Given the size and the inconsistency and interactions of the effects across study designs, a logical deduction would be that the effects of subjective factors within experimentation targeting objective performance measures are difficult, if not impossible, to estimate. This calls for a critical reflection of existing knowledge production via experimentation with differing input charts in visualization research, for instance of studies such as %Papers in which the charts used present potentially problematic combinations of different design parameters:
\citep{andry2021interpreting,10.1145/3411764.3445637,ajani2021declutter,li2014chart}.
When multiple properties of a chart are experimentally manipulated at the same time, it is difficult to attribute resulting effects to just one single dimension.

\subsection{Implications for data visualizations and their experimental evaluation}

The most immediate takeaway for understanding how people relate to charts is that the interaction between elements of charts is vitally important~\citep{delacuesta2022improving}.
When multiple properties of charts differ, it's not possible to simply decompose that effect as the sum of the effects for each individual element.
As we have shown in~\autoref{fig:paired_by_background}, effects differ substantially based on the background properties.
This is particularly a problem when some dimensions (such as source annotation) are more subtle than other elements (like bar orientation).
This highlights the need for larger-scale factorial experiments that can intentionally fix these additional confounding factors in experiments.
Such experiments provide the ability to unpack these diverse interaction effects and study them intentionally~\citep{egami2018causal}.

This is particularly problematic when using real-world charts, which differ on a large number of dimensions.
It is not possible, then, to clearly attribute differences between (for example) different orientations if the underlying data differs between charts or if other properties differ, even if these are simply aesthetic ones.
It is also particularly problematic based on the scope of chart elements that are to be modified in a study. If chart elements vary greatly in their subtlety, then this should be incorporated into the study design by ensuring that sample sizes of different combinations of factors are sufficient to fix background characteristics, as we have done in \autoref{fig:paired_by_background}.
\citet{delacuesta2022improving} provides further suggestions regarding experimental design in this setting.

We have found that the choice task was very good in clearly identifying the effects of clearly visible aesthetic factors. Still, to identify more subtle factors, it was more effective to use either the rating task or to focus on only the comparison between charts that were otherwise identical (except for the single dimension of interest). In particular, the choice task was very effective in revealing consistent (but weakly held) preferences, such as those on color and orientation.

Finally, we highlight two final points which should be considered in future experimental work on charts. First, we found qualitative validation of respondents' cognitive evaluation of charts crucial for unpacking their preferences. Hence, we recommend future studies to incorporate a qualitative component.
Second, we recommend being explicit about exactly which properties of charts are being varied in studies of chart design, as well as increased transparency in the precise task design. As this work shows, details of the task design can be crucial in how respondents evaluate charts.

In future work, we aim to explore the conceptual link between subjective experience and sensemaking. This includes, e.g., the experienced understandability of charts or the perceived cognitive effort of reading them and their relationship to \textit{objective} levels of complexity.

\section{Limitations}

\textit{Image size} was different between Study 1 and Study 2 due to the differences in task design (showing one image as opposed to two on a screen).

While \textit{participants} were recruited from the same general cohort, they were recruited from two different semesters of a university course for Study 1 and Study 2, respectively. They represent a WEIRD demographic~\citep{linxen2021weird} composed of young computer science students. It would further be interesting to explore potential differences in the experience of certain design parameters based on levels of expertise, as this has shown to impact the interpretation of charts~\citep{peebles2015expert}.

The \textit{sample of charts} used in the study was informed by real-world data. However, it still presents artificially created charts with very limited complexity. 

%Artifical charts -> contrasted by results from the real charts
%Participant sample not ideal because WEIRD computer science students

\section{Conclusion}

This work investigates the impact of study design targeting subjective experience in the context of visualization research. To do this, we study the impact of design changes in simple bar charts on perceived readability and trustworthiness.
We report on the results from a triangulation of methods to reflect on experimental designs and by grounding our results in a qualitative account of participants reasoning for decision making, which serves to contextualize and validate the studies. Our results suggest thoroughly considering the influence of differences in the salience of design changes on experimental results and carefully isolating individual elements in experimental chart evaluation, even if the changes are seemingly inconsequential.

\section*{Acknowledgements}
This work has been funded by the Vienna Science and Technology Fund (WWTF)[10.47379/ICT20065].
%This section will be added later as it does not comply with the anonymous review policy.
%We thank our participants for their time and contribution. We thank Simone Jochum and Alexander Seitner for their assistance.
%\input{supplementary}
%\subsection{Supplementary material}

\bibliographystyle{ACM-Reference-Format}
\bibliography{sample-base}

\end{document}